\documentclass[journal]{IEEEtran}
\IEEEoverridecommandlockouts
\usepackage[utf8]{inputenc}
\usepackage[english]{babel}
\usepackage{cite}
\usepackage{amsmath,amssymb,amsfonts}
\usepackage{graphicx}
\usepackage{subfigure}
\usepackage{epstopdf}
\usepackage{textcomp}
\usepackage{verbatim}
\usepackage{xcolor}
 \usepackage{url}
\usepackage{algorithm}
\usepackage{multirow}
\usepackage{cancel}
\usepackage{xargs}
\usepackage{amsthm}

\ifCLASSINFOpdf
\else
\fi

\begin{document}
\bibliographystyle{IEEEtran}

\title{Covert Communication for Untrusted UAV-Assisted Wireless Systems}

\author{Chan Gao, Linying Tian, Dong Zheng
\thanks{Chan Gao, Linying Tian and Dong Zheng are with the National Engineering Research Center for Secured Wireless, School of Cybersecurity, Xi'an University of Posts and Telecommunications, Xi'an 710121, China (e-mail:gaochan001@163.com; tianlinying129@163.com; zhengdong@xupt.edu.cn, corresponding author: Chan Gao). }
}

\maketitle
\begin{abstract}
Wireless systems are of paramount importance for providing ubiquitous data transmission for smart cities. However, due to the broadcasting and openness of wireless channels, such systems face potential security challenges. UAV-assisted covert communication is a supporting technology for improving covert performances and has become a hot issue in the research of wireless communication security. This paper investigates the performance of joint covert and security communication in a tow-hop UAV-assisted wireless system, where a source transmits the covert message to a destination with the help of an untrusted UAV. We first design a transmission scheme such that use UAVs to assist in covert communications while ensuring the security of covert messages. Then, we develop a theoretical model to derive the expressions for the detection error probability of the warden and the covert and security rate, and the maximum covert and security rate is optimized by power control under a given covertness and security requirements.
Finally, numerical results are provided to illustrate our theoretical analysis and the performance of covert and security communication in such systems.
\end{abstract}

\IEEEpeerreviewmaketitle

\section{Introduction}
With the continuous advancement of technology, many new wireless communication technologies have emerged, which use wireless media to transmit information and have changed our way of life to a great extent. Although wireless communication technology has made great progress and development, there are still some problems and challenges, such as the broadcasting nature of the wireless channel itself, limited spectrum resources for wireless communication, etc. In order to solve these problems, there is an urgent need for an effective method to protect our communication from being detected.

In recent years, the Internet of Things (IoT) technology has been applied to wireless communication, which is a technology that connects various physical devices, sensors, software, and networks to achieve intelligent interoperability and interoperability through the Internet, which allows us to realise the integration of the physical world and the digital world. Through IoT technology, we can carry out data interaction, remote control and automated operation between devices, thus providing smarter and more convenient solutions for people's life and work \cite{6}. The Internet of Things (IoT)-centric concepts like augmented reality, high-resolution video streaming, self-driven cars, smart environment, e-health care, etc. have a ubiquitous presence now. Whether it's for smart predictive maintenance for manufacturing, WI-FI network services for new energy-efficient access point systems, or 5G wireless systems and next-generation smart systems, IoT technology has a non-negligible involvement \cite{7,8,9,10}. However, due to the fact that a large amount of information, especially secret information, is transmitted through IoT, there are certain security issues that may lead to problems such as data loss, stealing or tampering of information.

The main existing communication protection methods are cryptography and physical layer security (PLS). Cryptographic techniques are used to protect the privacy and integrity of the data by converting the communication data into ciphertext using cryptographic algorithms, and only the receiver with the decryption key can restore the ciphertext to plaintext. It is worth noting that cryptographic techniques have high computational requirements in some cases, especially for some complex encryption algorithms, the encryption and decryption processes may require a large amount of computational resources and time, thus there is a delay. In recent years, physical layer security has been studied and indicated as a possible way to emancipate networks from classical, complexity based, security approaches\cite{10044975}. Physical layer security technology refers to a series of technical measures taken on the physical layer of a communication system (i.e., the physical medium of signal transmission) to protect data transmission and information security in the communication process. Unlike traditional network layer and application layer security technologies, it works directly on the physical transmission medium, providing an additional layer of security protection that can enhance the security of the entire communication system. Physical layer security techniques are computationally light as compared to cryptography and are widely used for data security. And an obvious problem with both cryptographic and physical layer security techniques is that it only provides protection for the data, but does not hide the communication itself.

In particular, covert communication technology is a very promising technology with great potential for protecting communication security and privacy, and has important applications in many fields such as military, intelligence, and Internet of Things. The goal of covert communication is to hide the communication behaviour while the user is communicating in order to better protect the user's privacy and prevent the eavesdropping of secret information.
\subsection{Related Works}
With the rapid development of wireless communication, more and more sensitive information, such as personal privacy and financial data, is transmitted over wireless channels. Naturally, people put forward higher requirements for the security of wireless communication\cite{9790826}. It will use some hiding or camouflage methods to avoid detection when transmitting information to protect the security and privacy of communication, or secret communication in some special environments. Currently, research on covert communication involves two main systems: single-hop and two-hop systems. In the single-hop system, the transmitter and receiver of a message communicate only once in a direct communication without the assistance of others. The two-hop system is more complex than the single-hop system, in which the transmitter and receiver of a message communicate through an intermediate node, which usually uses encryption and masking techniques so that the message is not intercepted or interpreted by a third side in the transmission process.

In single-hop systems, it was shown in \cite{6584948} that $\mathcal{O}(\sqrt{n})$ bits can be sent from the transmitter to the receiver in n channel use. Single-hop systems are mainly considered in a typical three-node model consisting of a legitimate transmitter-receiver pair (Alice and Bob) and a warden (Willie). \cite{g1, g2, g3, g4} conducted studies for different kinds of jamming, where \cite{g1} deployed an additional cognitive jammer and studied secret communication with the help of a jammer, the one used was able to decide whether to transmit a jamming signal or not, based on the results of its perception. The jammer used in \cite{g2} is the same as the one in the above article, but with the difference that the jammer is able to generate intermittent artificial noise, which is more conducive to the concealment of secret messages. In contrast, in \cite{g3}, the transmitter is made to be equipped with two antennas, one for transmitting the secret message and the other for transmitting a jamming signal to confuse the detecting party. A covert communication scheme assisted by multiple antenna jammers is also proposed in \cite{g4}, with corresponding antennas transmitting jammers for multiple randomly distributed detection parties. \cite{p1, p2, p3} are studied for power control, and in \cite{p1} a channel inversion power control scheme is proposed, which allows the transmitter to communicate with the receiver without having to emit guided signals due to channel reciprocity, which facilitates the concealment of the transmitter in covert communications. The situation in \cite{p2} is that the covert user is shielded by a public user with truncated channel inversion power control that acts as a random transmit power jammer to interfere with the warden's detection. An adaptive power control strategy with partial channel state information is proposed in \cite{p3}, which reduces the outage probability while introducing more uncertainty to the warden. Literature \cite{x1, x2, x3} addresses different channel choices, \cite{x1} chose a slow fading channel and studied covert communication between a pair of legal transmitters and receivers and detectors. \cite{x2} investigated finite block length covert communication with randomly selected  channel uses, where the transmitter randomly selects the channel uses for transmission in an additive gaussian white noise environment\cite{x3}. Noise-free causal feedback in covert communication over a two-user gaussian multiple-access channel is introduced and its fundamental limits are analysed. All these works optimise the performance of the covert system, i.e., the probability of detection error and the covert rate.

In a two-hop system, the communicating sides are connected to each other through relay nodes for the purpose of communication, where the selection of the nodes and the establishment of the communication paths are critical and need to be considered for a number of factors such as covertness, reliability, and efficiency. Covert communication with relay selection in relay networks is studied in \cite{d1}, the scenario considered is that while forwarding a source message, the selected relay takes advantage of the opportunity to secretly transmit its own message to the receiver. Covert wireless communication in multi-antenna relay networks is studied in \cite{d2}, where the relay also transmits its own covert message to the receiver at the destination while assisting in the delivery of the source's message, and importantly, at this point the transmitter acts as a detector to detect this covert transmission. And the basic covert rate performance of a wireless relay system consisting of a transmitter and a receiver, a relay and a detector is investigated in \cite{d3}, where the relay can be switched in full-duplex mode or half-duplex mode depending on the channel state of the self-interfering channel, and finally the covert rate performance of the system is investigated in various scenarios.

In recent years, in order to improve the covertness and transmission distance of covert communications, technologies such as UAVs and reflective surfaces have been added, and the introduction of these technologies can increase the complexity and resistance of covert communication systems, and improve the security and covertness of hidden information. At the same time, the combination of technologies such as UAVs and reflective surfaces can also be applied in specific environments, such as military intelligence collection, security monitoring and other fields, to further enhance the functionality and utility of covert communication systems. UAVs can act as transmitters to send secret messages \cite{w1, w2}, as jammers to send jamming signals to confuse detection \cite{w3}, and as relays to assist in covert transmissions between transmitters and receivers \cite{w4}. The mobility of UAVs makes communication links difficult to detect and jam, while also being able to extend the transmission distance of covert communications. Reflective surface technology can alter signal propagation paths through the use of surface reflections, making the source and destination of communication signals more difficult to detect, allowing for more flexible and undetectable covert communications \cite{f1, f2}.
\subsection{Motivation and Contributions}
However, all of the above studies on UAVs are about the situation where UAVs are completely trustworthy, which is less common in practical applications, where UAVs usually eavesdrop on the signals to be relayed when they act as relays, which is a potential threat to our entire covert communication system. Therefore, when combining UAV technology for covert communication system research, the situation that UAVs are not fully trustworthy needs to be taken into account.

In order to conduct our study clearly, the rest of the paper is structured as follows. Section II is the system model, which is divided into two parts: covert communication scenarios and system hypothesis testing. Section III is the performance constraints analysis of the proposed system in this paper, mainly two kinds of covert constraints and security constraints, and covert rate maximisation analysis. Section IV presents the numerical results, including model validation and theoretical analysis of the security covert performance. The last section concludes the paper.

\section{SYSTEM MODELS}
In this section, we firstly analyse the scenario in which this covert communication takes place, introduce the channel and noise used in the communication, then the detector performs a binary hypothesis test based on the received signal to get the required probability of detection error, and finally analyse the covert requirements of the system.

\begin{figure}[t]
\centering
\includegraphics[width=0.47\textwidth]{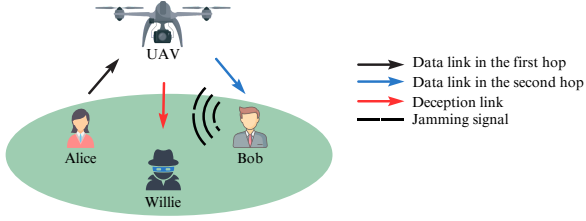}\\
\DeclareGraphicsExtensions.
\caption{Covert communication scenario.}
\label{Fig.1}
\end{figure}

\subsection{Covert Communication Scenario}

As shown in Fig.1, we consider a covert communication scenario with a Unmanned Aerial Vehicle(UAV) as a relay, including a UAV relay hovering in the air and the transmitter(Alice), the full-duplex receiver(Bob), and the warden(Willie) on the ground. Without loss of generality, we define the horizontal positions of Alice, UAV, Bob, and Willie as $L_{a}=[x_{a},y_{a}]$, $L_{u}=[x_{u},y_{u}]$, $L_{b}=[x_{b},y_{b}]$ and $L_{w}=[x_{w},y_{w}]$, respectively.

The scenario altogether includes two kinds of communication processes, i.e., the information transmission process and the detection process. Among them, the information transmission process includes the first period from Alice to UAV and the second period from UAV to Bob, while the detection process is the detection of UAV by Willie. The information transmission process is as follows: firstly, Alice sends covert information $\mathbf{x}_a(i)$ to UAV with transmit power $P_{a}$, where $i$ represents the index used by $N$ channels, $\mathbf{x}_a(i) \sim CN(0,1)$, and UAV receives the information, amplifies and forwards it, and relays the amplified signal $\mathbf{x}_u(i)$ to Bob with power $P_{u}$. What's worth noting is that the UAV relay devices will eavesdrop, so the amount of information for Bob, the receiver, will be reduced. The detection process is as follows: Willie determines whether the UAV is forwarding or not based on the signals he receives, and in order to disturb the detecting partner, we use one of the ports of the full-duplex receiver to release a jamming signal for the purpose of transmitting the covert message.

It has been proved that $\mathcal{O}(\sqrt{n})$ bits of information can be reliably and covertly transmitted to a legitimate receiver in n-channel use, as n tends to infinity.

We assume that the UAV relay hovers at altitude $H$, $H$ is not fixed, and during the communication process, the UAV provides a LOS link to the ground device, therefore, the channel gain from the UAV to the ground device can be denoted as:
\begin{flalign}
h_{ui}=\sqrt{\frac{\beta}{\parallel \mathbf{L_{u}} - \mathbf{L_{i}} \parallel^2+H^2}},\,\,\,i=a,w,b,
\end{flalign}
where $\beta$ is the channel power gain at a reference distance of 1$m$. The channel from Bob to Willie and Bob's own channel can be defined as $h_{bw}$ and $h_{bb}$, respectively, and they both obey quasi-static Rayleigh fading, and the probability density function of $h_{bb}$ can be denoted as $f(x)=e^{-x}$, We assume that the UAV knows $h_{ua}$, $h_{ub}$, $h_{uw}$, Bob knows only $h_{ub}$ and Willie knows only $h_{uw}$.

In the first stage of covert information transmission, if Alice transmitted a covert message, the received signal from the UAV is as below:
\begin{flalign}
 y_{u}(i)=\sqrt{P_{a}}|h_{ua}|\mathbf{x}_a(i)+\sqrt{P_{J}}|h_{ub}|\mathbf{x}_J(i)+\mathbf{n}_u(i),
\end{flalign}
where $P_{a}$ is the power of the Alice transmitting the signal $\mathbf{x}_a(i)$, $P_{J}$ is the power of the receiver transmitting the jamming signal, and $\mathbf{n}_u(i)$ is the additive Gaussian white noise with $\sigma_{u}^2$ as the variance at the UAV, i.e. , $\mathbf{n}_u(i)\sim(0,\sigma_{u}^2)$. In the second stage of covert message transmission, the UAV forwards a linearly scaled version of the received signal to the receiver Bob, so the forwarded signal of the UAV is:
\begin{flalign}
\mathbf{x}_u(i)=G[\mathbf{y}_u(i)],
\end{flalign}
which is a linearly scaled version of the received signal with $G$ as a scalar, and the value of $G$ is chosen to satisfy $E[\mathbf{x}_u(i)\mathbf{x}_u(i)^\dag]=1$ in order to ensure the power constraints at Bob, thus obtaining the value of $G$:
\begin{flalign}
G=\frac{1}{\sqrt{P_{a}|h_{ua}|^{2}+P_{J}|h_{ub}|^{2}+\sigma_{u}^2}}.
\end{flalign}

\subsection{System Hypothesis Test}
In an attempt to detect whether the UAV has made a forwarding, Willie uses a binary hypothesis test, i.e., based on his own observed signals, he chooses one of the null hypothesis $H_{0}$ and alternative hypothesis $H_{1}$. The null hypothesis $H_{0}$ indicates that the UAV has not made a forwarding, i.e., that the transmitter, Alice, has not transmitted a covert message, and the alternative hypothesis $H_{1}$ is the contrary.

With the two hypotheses, Willie's observed signal can be given as:
\begin{flalign}
&\mathbf{y}_w(i)= \nonumber\\
&\left\{
\begin{aligned}
&\sqrt{P_{J}} |h_{bw}|  \mathbf{x}_J(i) +  \mathbf{n}_w(i), & \text{if}\,H_{0}\,\text{is~true} \\
&\sqrt{P_{u}} |h_{uw}|  \mathbf{x}_u(i) + \sqrt{P_{J}} |h_{bw}|  \mathbf{x}_J(i) +  \mathbf{n}_w(i),&\text{if}\,H_{1}\,\text{is~true}
\end{aligned}
\right.
\end{flalign}
where $P_{u}$ is the power used by the UAV to forward the covert message $ \mathbf{x}_u(i)$, and $ \mathbf{n}_w(i)$ is the additive Gaussian white noise at Willie with $0$ as the mean and $\sigma_{w}^2$ as the variance.

According to Leberger's dominated convergence theorem, Willie's test statistic, i.e., the average received power, as $n$ tends to infinity can be expressed as:
\begin{flalign}
T(y_{w}) &=
\frac{1}{N} \sum_{i=1}^{N}{|\mathbf{y}_w(i)|}^{2}\nonumber\\
&=\left\{
\begin{aligned}
&P_{J}|h_{bw}|^{2}+\sigma_{w}^{2}, &
\text{if}\,H_{0}\,\text{is~true} \\
&P_{u}|h_{uw}|^{2}+P_{J}|h_{bw}|^{2}+\sigma_{w}^{2}.&
\text{if}\,H_{1}\,\text{is~true}
\end{aligned}
\right.
\end{flalign}

In order to minimum the total test error probability at Willie, the optimal decision rule according to the Nyman-Pearson criterion is:

\begin{flalign}\label{T(y_{w})}
T(y_{w})\underset{D_{0}}
{\overset{D_{1}}\gtrless}\gamma,
\end{flalign}
where $D_{0}$ and $D_{1}$ denote the decisions made by Willie under the assumptions that $H_{0}$ and $H_{1}$ hold, respectively, and $\gamma$ is the detection threshold.

Therefore, we can define the probability of false alarm and the probability of missed detection as ${\mathbb P}_{FA}$, ${\mathbb P}_{MD}$, respectively, and they can be written as: $\mathbb{P}_{FA}=P(D_{1}|H_{0})$, $\mathbb{P}_{MD}=P(D_{0}|H_{1})$.

Then the total test error probability at Willie stands for:
\begin{align}
    \zeta = {{\mathbb P}_{FA}} + {{\mathbb P}_{MD}}.
\end{align}

Standing in Willie's point of mind, it wants to find an optimal detection threshold $\gamma^*$ such that the total detection error probability is minimum, i.e., $\zeta^* = \zeta(\gamma^*)$. For Alice, UAV and Bob, on the other hand, they want the minimum test error probability at Willie to always be greater than a specific value $1-\varepsilon$, where $\varepsilon$ is an arbitrarily small positive number, so that we consider the requirements for covert communication are met.

\section{Performance Constraint}
In this subsection, we detailed the two constraints in the system, namely, the covert constraint and the security constraint, and summarised the maximum covert transmission rate that can be achieved by the system under the two constraints.
\subsection{Covert Constraint}
Based on $\mathbb{P}_{FA}=P(D_{1}|H_{0})$, $\mathbb{P}_{MD}=P(D_{0}|H_{1})$, (6) and (7), we can get the probability of false alarm and probability of missed detection for Willie:
\begin{flalign}
\mathbb{P}_{FA} &=  P(P_{J}{|h_{bw}|}^{2} + \sigma_{w}^{2} > \gamma)\nonumber\\
&= P \left({|h_{bw}|}^{2} > \frac{\gamma - \sigma_{w}^{2}}{P_{J}}\right) \nonumber\\
&=\left\{
\begin{aligned}
&\text{exp}\left(\frac{\sigma_{w}^{2} - \gamma}{P_{J}}\right), &\text{if} \,\gamma \geq \sigma_{w}^{2}\\
&1,&\text{otherwise}
\end{aligned}
\right.
\end{flalign}
and
\begin{flalign}
\mathbb{P}_{MD} &=  P(P_{u}{|h_{uw}|}^{2}+P_{J}{|h_{bw}|}^{2} + \sigma_{w}^{2} < \gamma)\nonumber\\
&= P \left({|h_{bw}|}^{2} < \frac{\gamma - \sigma_{w}^{2}-P_{u}{|h_{uw}|}^{2}}{P_{J}}\right) \nonumber\\
&= P \left({|h_{bw}|}^{2} < \frac{\gamma - \sigma_{w}^{2}-\frac{P_{u}\beta}{d_{w}^{2}+H^2}}{P_{J}}\right) \nonumber\\
&=\left\{
\begin{aligned}
&1-\text{exp}(\tau), &\text{if} \,\gamma \geq \sigma_{w}^{2}+\frac{P_{u}\beta}{d_{w}^{2}+H^2}\\
&0,&\text{otherwise}
\end{aligned}
\right.
\end{flalign}
where $\tau=(\sigma_{w}^{2} + P_{u}\beta/(d_{w}^{2}+H^2)-\gamma)/{P_{J}}$.

By associating (8), (9) and (10), we can derive the optimal detection threshold $\gamma^{*}$ and the corresponding minimum detection error probability $\zeta^{*}$ at Willie.

\text{Theorem 1}: The optimal detection threshold at Willie is $\gamma^{*}=\sigma_{w}^{2}+P_{u}\beta/(d_{w}^{2}+H^2)$, and the corresponding minimum detection error probability is $\zeta^{*}
=\text{exp}(-P_{u}\beta/P_{J}(d_{w}^{2}+H^{2}))$.

\text{Proof}: The total detection error probability at Willie is given according to the equations (8), (9), and (10),
\begin{flalign}
&\zeta= &  \\
&\left\{
\begin{aligned}
&1,\quad \text{if} \,\gamma\leq \sigma_{w}^{2}\nonumber\\
&\text{exp}\left(\frac{\sigma_{w}^{2}-\gamma}{P_{J}}\right), \quad \text{if} \, \sigma_{w}^{2}< \gamma \leq  \sigma_{w}^{2}+ \frac{P_{u}\beta}{d_{w}^{2}+H^{2}}\nonumber\\
&1-\text{exp}(\tau)+\text{exp}\left(\frac{\sigma_{w}^{2}-\gamma}{P_{J}}\right), \quad \,\text{if} \, \gamma>\sigma_{w}^{2}+ \frac{P_{u}\beta}{d_{w}^{2}+H^{2}} \nonumber
\end{aligned}
\right.
\end{flalign}

Analysing the above equation, we can easily see that when $\gamma\leq\sigma_{w}^{2}$, $\zeta$ is a constant value; when $\gamma\in(\sigma_{w}^{2},\sigma_{w}^{2}+P_{u}\beta/(d_{w}^{2}+H^2))$, $\zeta$ decreases with increasing $\gamma$; but when $\gamma$ is larger than $\sigma_{w}^{2}+P_{u}\beta/(d_{w}^{2}+H^2)$, the amplitude of change of $\zeta$ with $\gamma$ is uncertain. Therefore, when $\gamma$ is greater than $\sigma_{w}^{2}+P_{u}\beta/(d_{w}^{2}+H^2)$, we derive $\zeta$. The first order derivative of $\zeta$ with respect to $\gamma$ becomes:

\begin{flalign}\label{eq:parzata}
\frac{\partial\zeta}{\partial\gamma}=
\frac{\text{exp}\left(\frac{\sigma_{w}^{2}+\frac{P_{u}\beta}
{d_{w}^{2}+H^{2}}-\gamma}{P_{J}}\right)
-\text{exp}\left(\frac{\sigma_{w}^{2}-\gamma}{P_{J}}\right)}
{P_{J}}.
\end{flalign}

we cannot clearly determine the trend of the first-order derivative from the expression, so we derive it again to obtain the expression for the second-order derivative as follows:

\begin{flalign}\label{eq:parzata}
\frac{\partial^{2}\zeta}{\partial\gamma^{2}}=
\frac{\text{exp}\left(\frac{\sigma_{w}^{2}-\gamma}{P_{J}}\right)-
\text{exp}\left(\frac{\sigma_{w}^{2}+\frac{P_{u}\beta}{d_{w}^{2}+H^2}-\gamma}{P_{J}}\right)}
{P_{J}^{2}},
\end{flalign}
according to the above equation we can easily get the second order derivative monotonically increasing, then, we make it equal to $0$, we can get:

\begin{flalign}
\gamma^{1}=\sigma_{u}^{2}+\frac{P_{u}\beta}{2(d_{w}^{2}+H^2)},
\end{flalign}
(false)

since the second-order derivative is $0$ at $1$ and the second-order derivative is monotonically increasing, what we can know is that $\gamma^{1}$ is the point of minimum of the first-order derivative function, and the minimum of the first-order derivative can be obtained by substituting (14) into (12) as below:

\begin{flalign}\label{eq:parzata}
\frac{\partial\zeta}{\partial\gamma}_{\text{min}}=
\frac{P_{u}\beta}{(d_{w}^{2}+H^2)P_{J}^{2}},
\end{flalign}
therefore, we can clearly see that the minimum value of the first-order derivative is greater than $0$. Therefore, the first-order derivative is greater than $0$, the original function is monotonically increasing, so when $\gamma>\sigma_{w}^{2}+P_{u} \beta/(d_{w}^{2}+H^2)$, $\frac{\partial\zeta}{\partial\gamma}$ is an increasing function with respect to $\gamma$, and hence, the optimal detection threshold and the minimum detection error probability for Willie is as Theorem 1.

Depending on Theorem 1, we can acquire the covert constraint $\zeta^{*} \geq 1-\varepsilon$. Based on this, we can derive the range of hover height: $H\geq\sqrt{-P_{u}\beta/P_{J}\ln(1-\varepsilon)-d_{w}^{2}}$.

\subsection{Security Constraint}

In addition to the covert constraint, we will define a security constraint that implies the security of the transmission between the UAV and Bob, and break if our security requirements are not met.

We determine the difference between the channel capacity $C_{b}$ at the receiving point Bob and the signal capacity $C_{u}$ stolen by the UAV as the secure transmission rate $C_{s}$ and predetermine a secure rate threshold $R_{s}$. When the secure transmission rate $C_{s}$ is smaller than the threshold $R_{s}$ that we set, it means that the security rate that we require has not been reached, and that the UAV steals too much of the signal capacity, and that the transmission should be immediately interrupted.

On the basis of the previous analysis we know that in a situation where Alice has transmitted a covert message to the UAV, the reception of the UAV is (2), hence, its received signal-to-noise ratio $\gamma_{u}$ can be indicated as:

\begin{flalign}
\gamma_{u}&=\frac{P_{a}|h_{ua}|^{2}}
{P_{J}|h_{ub}|^{2}+\sigma_{u}^{2}}\nonumber\\
&=\frac{P_{a}\beta(d_{b}^{2}+H^{2})}
{(d_{a}^{2}+H^{2})(P_{J}\beta+\sigma_{u}^{2}(d_{b}^{2}+H^{2}))},
\end{flalign}
the channel capacity $C_{u}$ can be shown as:
\begin{flalign}
C_{u}=&\log_{2}(1+\gamma_{u})\nonumber\\
&=\log_{2}\left(1+\frac{P_{a}\beta(d_{b}^{2}+H^{2})}
{(d_{a}^{2}+H^{2})(P_{J}\beta+\sigma_{u}^{2}(d_{b}^{2}+H^{2}))}\right).
\end{flalign}

And the received signal at Bob is:

\begin{flalign}
\mathbf{y}_b(i)&=\sqrt{P_{u}}|h_{ub}|\mathbf{x}_u(i)+\sqrt{P_{J}}|h_{bb}|\mathbf{x}_J(i)+\mathbf{n}_b(i)\nonumber\\
&=\sqrt{P_{u}}|h_{ub}|\mathbf{x}_u(i)+\mathbf{n}_b(i),
\end{flalign}
the above process allows filtering the interference component of the received signal because the interference signal is generated by Bob itself, which allows self-interference cancellation and omission of interference terms.

And due to the fact that (3), (4),
therefore, its received signal can be rewritten as:

\begin{flalign}
\mathbf{y}_b(i)=&\sqrt{P_{u}}|h_{ub}|G[\sqrt{P_{a}}|h_{ua}|\mathbf{x}_a(i)+\sqrt{P_{J}}|h_{ub}|\mathbf{x}_J(i)\nonumber\\
&+\mathbf{n}_u(i)]+\mathbf{n}_b(i)\nonumber\\
=&\sqrt{P_{u}}\sqrt{P_{a}}|h_{ub}||h_{ua}|G\mathbf{x}_a(i)+\sqrt{P_{u}}|h_{ub}|G\mathbf{n}_u(i)\nonumber\\
&+\mathbf{n}_b(i).
\end{flalign}

Hence, its received signal-to-noise ratio $\gamma_{b}$ can be represented as:

\begin{flalign}
\gamma_{b}&=\frac{P_{u}P_{a}|h_{ub}|^{2}|h_{ua}|^{2}G^{2}}
{P_{u}|h_{ub}|^{2}G^{2}\sigma_{u}^{2}+\sigma_{b}^{2}}\nonumber\\
&=\frac{P_{u}P_{a}|h_{ub}|^{2}|h_{ua}|^{2}}
{(P_{u}\sigma_{u}^{2}+P_{J}\sigma_{b}^{2})|h_{ub}|^{2}+P_{a}|h_{ua}|^{2}\sigma_{b}^{2}+\sigma_{b}^{2}\sigma_{u}^{2}}\nonumber\\
&=\frac{P_{u}P_{a}\beta^{2}}
{\sigma_{b}^{2}\varphi+\sigma_{b}^{2}H^{2}\phi+H^{2}q+p+\sigma_{b}^{2}\sigma_{u}^{2}H^{4}},
\end{flalign}
and channel capacity $C_{b}$ can be expressed as:
\begin{flalign}
C_{b}=&\log_{2}(1+\gamma_{b})\nonumber\\
=&\log_{2}\left(1+\frac{P_{u}P_{a}\beta^{2}}
{\sigma_{b}^{2}\varphi+\sigma_{b}^{2}H^{2}\phi+H^{2}q+p+\sigma_{b}^{2}\sigma_{u}^{2}H^{4}}\right).
\end{flalign}

Then the secure transmission rate $C_{s}$ can be written as:

\begin{flalign}
C_{s}=&C_{b}-C_{u}\nonumber\\
=&\log_{2}\left(1+\frac{P_{u}P_{a}\beta^{2}}
{\sigma_{b}^{2}\varphi+\sigma_{b}^{2}H^{2}\phi+H^{2}q+p+\sigma_{b}^{2}\sigma_{u}^{2}H^{4}}\right)\nonumber\\
&-\log_{2}\left(1+\frac{P_{a}\beta(d_{b}^{2}+H^{2})}
{(d_{a}^{2}+H^{2})(P_{J}\beta+\sigma_{u}^{2}(d_{b}^{2}+H^{2}))}\right)\nonumber\\
=&\log_{2}\left(\frac{1+\frac{P_{u}P_{a}\beta^{2}}{p+\sigma_{b}^{2}\varphi+(q+\sigma_{b}^{2}\phi)H^{2}+\sigma_{b}^{2}\sigma_{u}^{2}H^{4}}}
{1+\frac{P_{a}\beta d_{b}^{2}+P_{a}\beta H^{2}}{\varphi+\phi H^{2}+\sigma_{u}^{2}H^{4}}}\right),
\end{flalign}
where
$$\varphi=P_{J}\beta d_{a}^{2}+\sigma_{u}^{2}d_{a}^{2}d_{b}^{2}$$ $$\phi=P_{J}\beta+\sigma_{u}^{2}(d_{a}^{2}+d_{b}^{2})$$ $$p=P_{u}\sigma_{u}^{2}d_{a}^{2}\beta+P_{a}\sigma_{b}^{4}\beta$$ $$q=P_{u}\sigma_{u}^{2}\beta+P_{a}\sigma_{b}^{2}\beta.$$

From the above analysis, we can see that interruption is required when the secure transmission rate $C_{s}$ is smaller than our pre-set secrecy rate $R_{s}$, and thus our security constraint requirement $C_{s}<R_{s}$ can be redrafted as:

\begin{flalign}
\log_{2} \left(\frac{1+\frac{P_{u}P_{a}\beta^{2}}{p+\sigma_{b}^{2}\varphi+(q+\sigma_{b}^{2}\phi)H^{2}+\sigma_{b}^{2}\sigma_{u}^{2}H^{4}}}
{1+\frac{P_{a}\beta d_{b}^{2}+P_{a}\beta H^{2}}{\varphi+\phi H^{2}+\sigma_{u}^{2}H^{4}}}\right) <R_{s},
\end{flalign}
combining with (22), we can obtain the range of hovering height $H$, $H\in(0,H^{'})$, where $H^{'}$ is the solution satisfying (23).

\subsection{Covert Rate Maximization}

Our ambition is to maximise the secure covert rate $R_{b}$, and $R_{b}=C_{b}$. In addition to Covert Constraint and Security Constraint, we also require that the transmit power $P_{a}$ and $P_{J}$ do not exceed the maximum transmit power $P_{max}$. So we can represent the optimisation problem for the system covert rate as:
\begin{align}\nonumber
{\text{maximize}} \quad &R_{b}\nonumber\\
\text{\textit{s.t.}} \quad
&\zeta^{*} \geq 1 - \varepsilon,\nonumber\\
&\sqrt{-P_{u}\beta/P_{J}\ln(1-\varepsilon)-d_{w}^{2}}\leq H<H^{'},\\
&\varepsilon\in(0,1),\\
&P_{u}\leq P_{max},\\
&P_{J}\leq P_{max},\\\nonumber
\end{align}
where the expression of $R_{b}$ as below:
\begin{flalign}
R_{b}
=&C_{b}\nonumber\\
=&\log_{2}\left(1+\frac{P_{u}P_{a}\beta^{2}}
{\sigma_{b}^{2}\varphi+\sigma_{b}^{2}H^{2}\phi+H^{2}q+p+\sigma_{b}^{2}\sigma_{u}^{2}H^{4}}\right).
\end{flalign}

\section{Numerical Results}
In this subsection, we use simulation tools to demonstrate our scheme with images that clearly show the variation of the detection error probability and the maximum covert rate of the system in order to evaluate the performance of the proposed transmission scheme.
\subsection{Model Validation}
In order to verify the efficiency of our proposed covert communication model in (11) and (22), we need to go ahead and analyse the detection error probability of the system and the final security and secrecy capacity.

The parameter settings used in the following text are: $\beta=10dB$, noise variance $\sigma_{i}^{2}=-20dB (i=u, b, w)$, and other special parameter settings are described separately.

In addition to this, in order to maximise the rate of covert, a comparison between the simulation results and the theoretical results was carried out, where the simulated detection error probability was calculated based on the average of $10^{5}$ independent simulations, and the value of the simulated detection error probability was equal to the ratio of the number of detection errors to the total number of detections.

Fig.2 analyses the amplitude of the total detection error probability $\zeta$ at Willie with respect to the detection threshold $\gamma$ for different UAV transmitting powers $P_{u}$. Where the transmit power of the interference signal $P_{J}=5W$, the horizontal distance difference between Willie and the UAV is $d_{w}^{2}=|L_{w}-L_{u}|^{2}=300m$, $L_{w}$ and $L_{u}$ are the horizontal positions of Willie and the UAV, respectively, while the hovering height of the UAV is $H^{2}=1000m$, and three sets of curves are obtained by three different values of the UAV transmit power $P_{u}=\{2, 3, 4\}W$, as shown as in Fig.2.

\begin{figure}[t]
\centering
\includegraphics[width=0.55\textwidth]{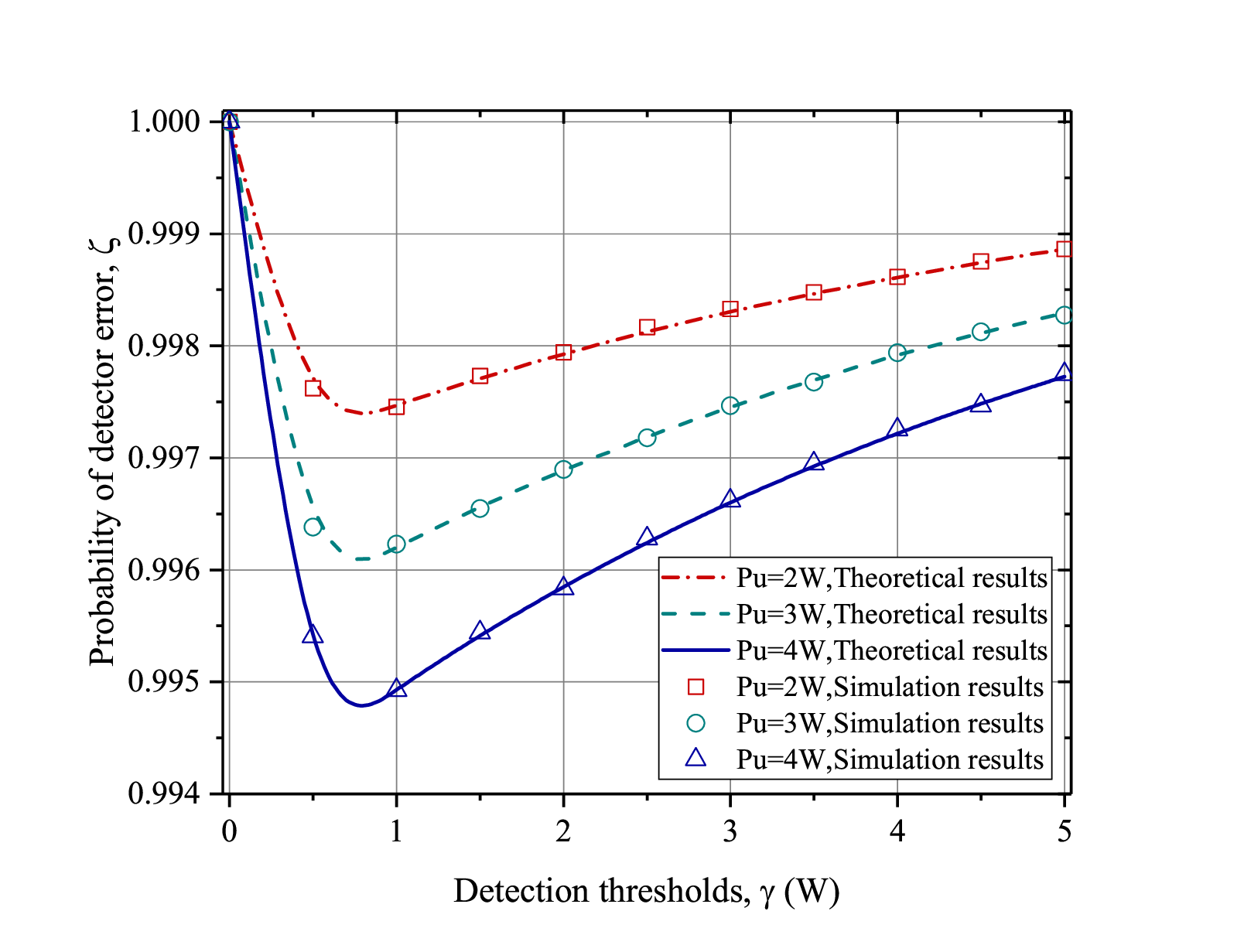}\\
\DeclareGraphicsExtensions.
\caption{The impact of detection threshold on detection error probability.}
\label{Fig.2}
\end{figure}

\begin{figure}[t]
\centering
\includegraphics[width=0.55\textwidth]{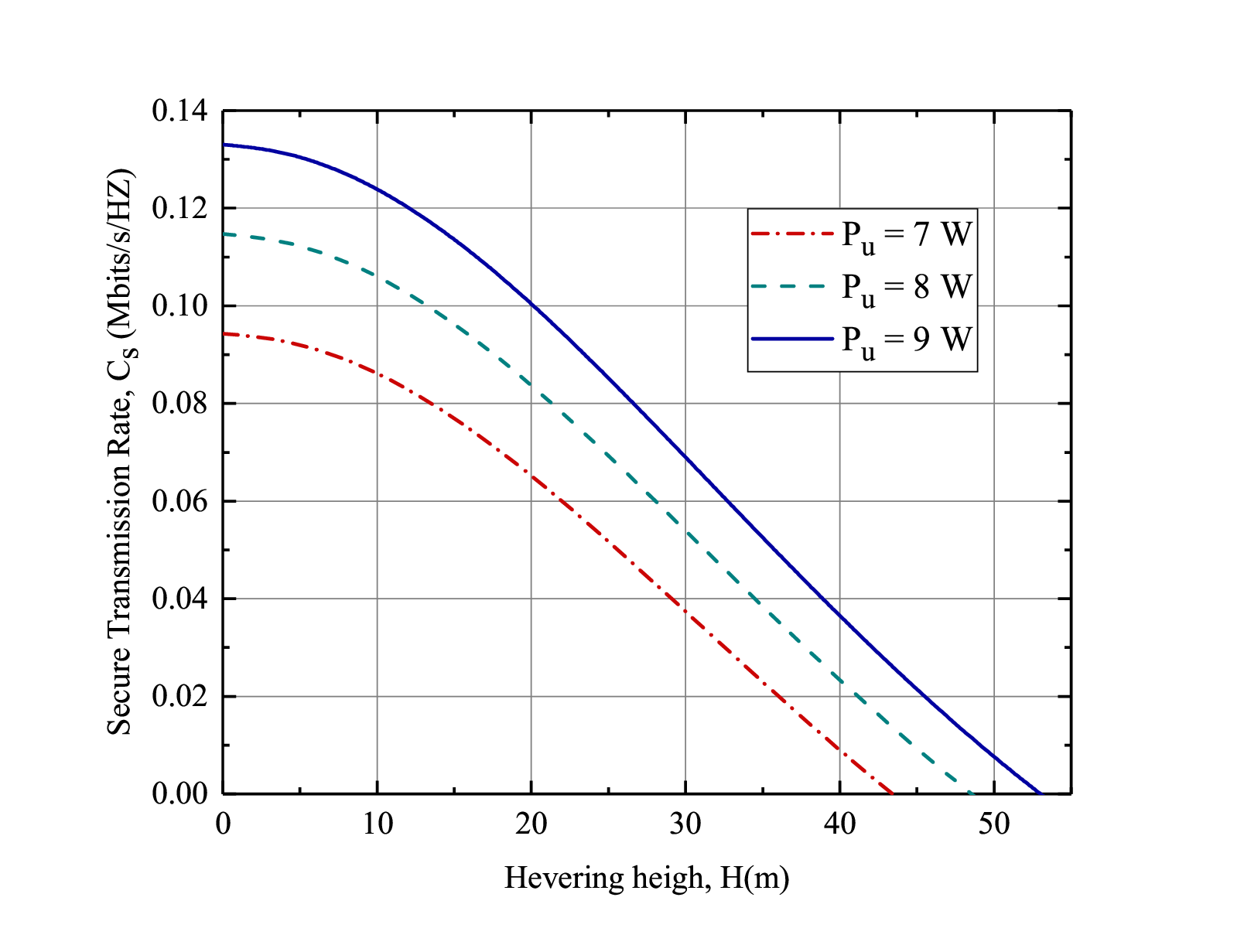}\\
\DeclareGraphicsExtensions.
\caption{The impact of hovering height on covert rate.}
\label{Fig.3}
\end{figure}

From Fig.2 we can see that the total detection error probability $\zeta$ first decreases as the detection threshold $\gamma$ increases, reaches a minimum point, and then increases as the detection threshold increases, so there is a minimum value of the total detection error probability, which is all obtained at about the detection threshold $\gamma= 0.5W$. And the minimum detection error probability decreases as the UAV transmitting power $P_{u}$ increases, this results from the fact that the signal strength Willie received increases and his detection becomes accurate, which leads to a reduction in the covert performance of the system. And we can observe from Fig.2 that the simulated detection error probability values are almost the same as the theoretical ones, which means that our theoretical scheme can predict the simulation results well.

\subsection{Theoretical Analysis of Security and Covert Performance}
Fig.3 is about the curve of secure transmission rate $C_{s}$, which breaks down how much the secure transmission rate of the system varies with the hovering height of the UAV under different UAV transmitting power $P_{u}$. We set the transmitting power of Alice as $P_{a}=2W$, the transmitting power of the jamming signal $P_{J}=10W$, the difference of the horizontal distance between Alice and UAV as $d_{a}=60m$, and the interpolation of the horizontal distance between Bob and UAV as $d_{b}=50m$, and three sets of curves as in Fig.3 are obtained by three different values of the UAV transmitting power $P_{u}$, $P_{u}={7, 8, 9}W$.

Fig.3 indicates that the secure transmission rate of the system $C_{s}$ decreases with the hovering height of the UAV and will decrease to $0$. As we can know from the previous analyses, although the higher the hovering height of the UAV, the worse the detection performance of Willie and the covert rate of the receiver will be decreased, so the hovering height of the UAV is not the higher the better. And the secure transmission rate increases when the UAV's transmitting power $P_{u}$ increases, which is because when the UAV's transmitting power increases, the receiver's covert rate also increases, which makes the secure transmission rate increase.

Fig.4 and Fig.5 both show curves on the maximum secure covert rate. Fig.4 analyses the increasing and decreasing properties of the maximum secure covert rate of the system $R_{b}$ with the hovering height of the UAV $H$ for different UAV transmitting powers $P_{u}$. The corresponding parameters are set as follows: the power used by Alice for transmitting the secret message is $P_{a}=1W$, the difference in horizontal distance between Alice and the UAV is $d_{a}=60m$, the difference in horizontal distance between Bob and the UAV is $d_{b}=50m$, and the transmitting power of the interfering signal is $P_{J}=5W$, and three sets of curves are obtained by three different values of the UAV's transmitting power $P_{u}=\{2, 3, 4\}W$, as shown as in Fig.4.

\begin{figure}[t]
\centering
\includegraphics[width=0.55\textwidth]{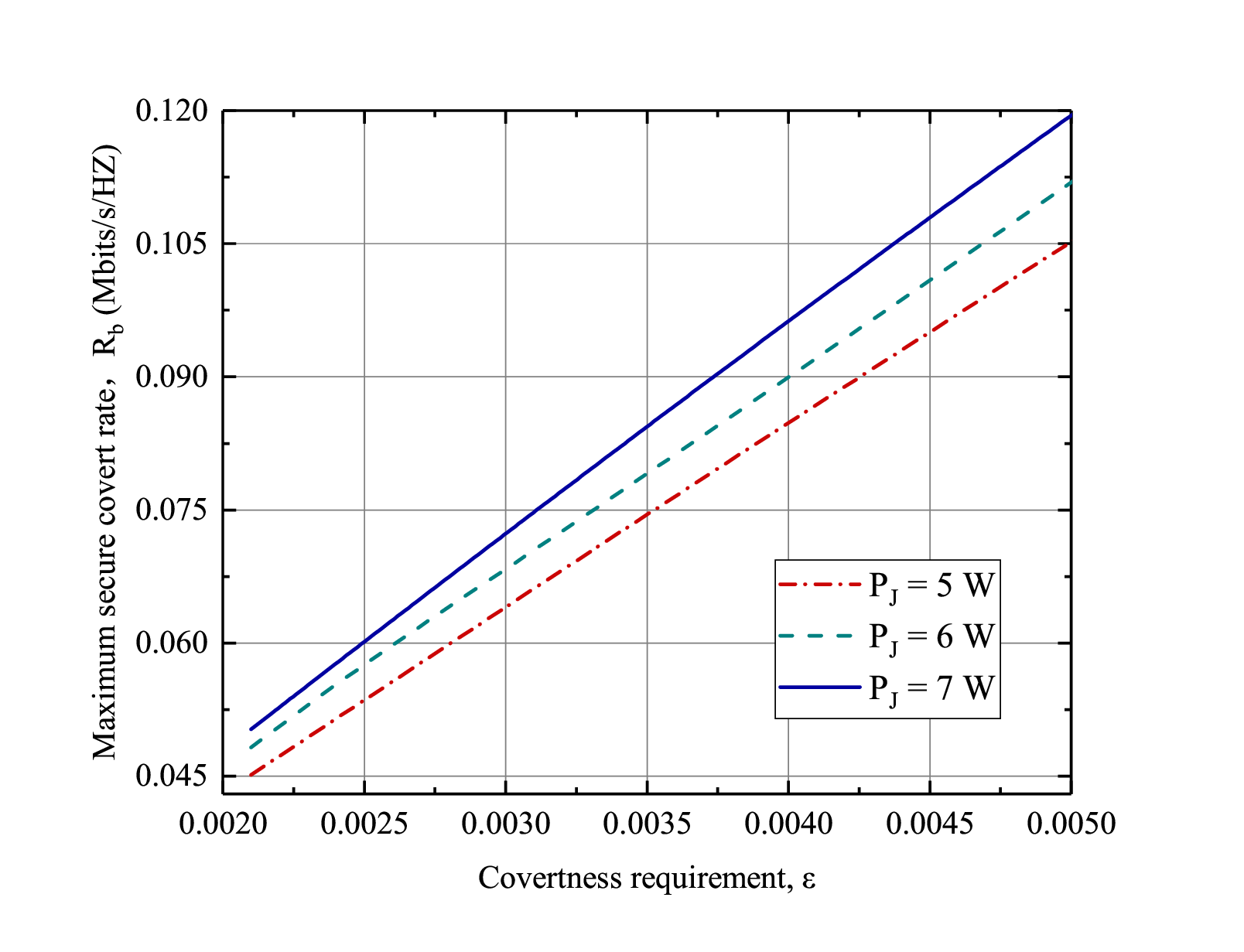}\\
\DeclareGraphicsExtensions.
\caption{The impact of Covertness requirement on covert rate.}
\label{Fig.4}
\end{figure}

Based on the curves shown in Fig.4, we can get the following conclusion that the maximum secure covert rate of the system $R_{b}$ decreases with the increase in the hovering height of the UAV $H$, which means that when the height of the UAV increases, the receiving rate of the receiver becomes smaller and the transmission performance of the system decreases. We can also observe that when the transmit power $P_{u}$ used by the UAV increases, the total covert transmission performance of the system also increases, this is due to the fact that at the same altitude, the UAV uses a greater transmit rate and the receive rate at the receiver increases.

Fig.5 illustrates the variation of the maximum secure covert rate $R_{b}$ with respect to the degree of covertness $\varepsilon$ when the emitted power of the jamming signal $P_{J}$ is different. In this case, the power used by Alice to transmit the secret message is $P_{a}=1W$, the difference in horizontal distance between Alice and the UAV is $d_{a}=60m$, and the difference in horizontal distance between Bob and the UAV is $d_{b}=50m$. According to the three different sets of UAV transmitting power $P_{u}$, different values of the minimum detection error probability are obtained, which results in different degrees of covertness $\varepsilon$. Three sets of curves are obtained under different interference powers $P_{J}$ as in Fig.5.

\begin{figure}[t]
\centering
\includegraphics[width=0.55\textwidth]{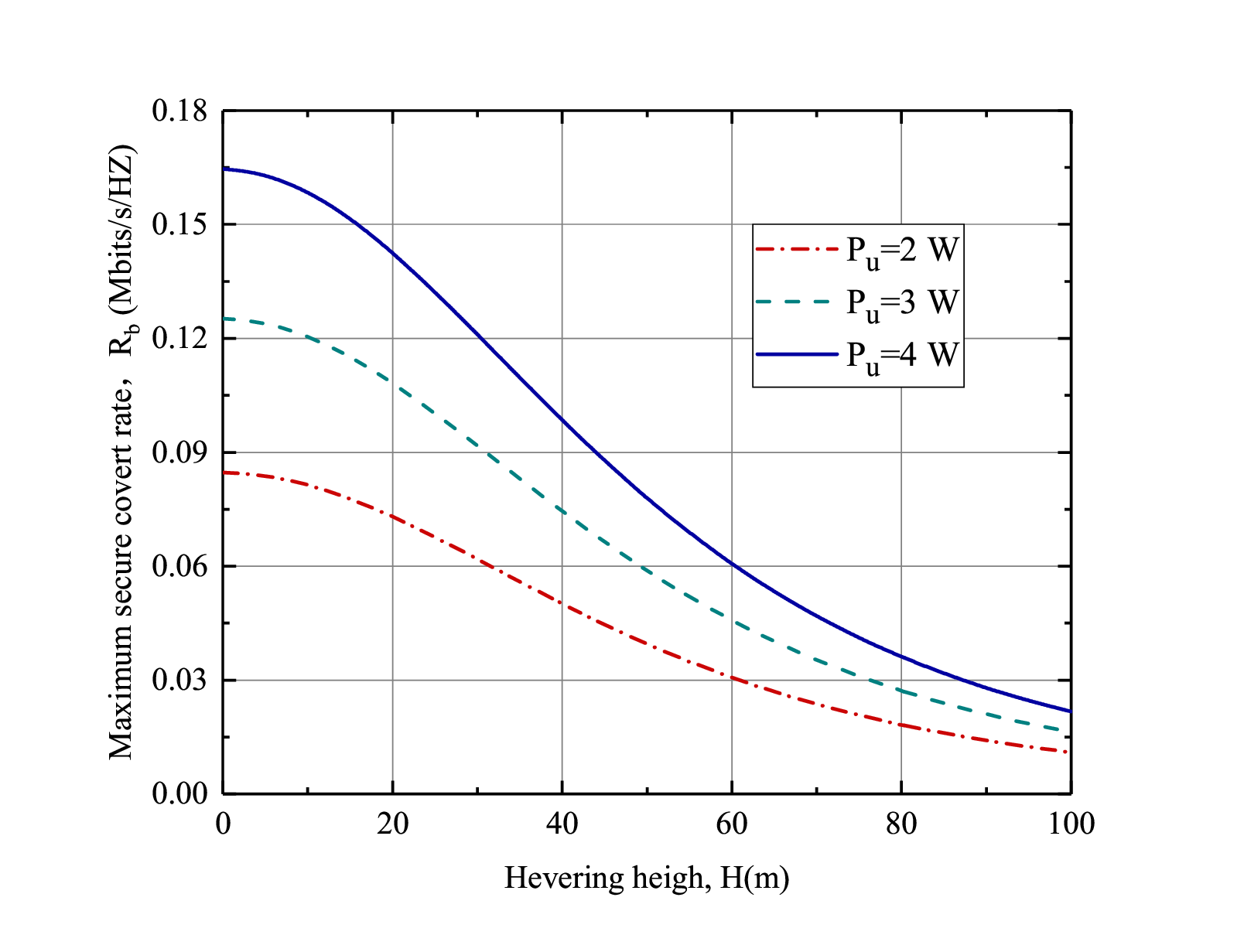}\\
\DeclareGraphicsExtensions.
\caption{The impact of covertness requirement on maximum covert capacity.}
\label{Fig.5}
\end{figure}

From Fig.5, it can be concluded that the maximum secure covert rate of the system $R_{b}$ increases with the increase of the covertness $\varepsilon$, the reason is that the larger the covertness, the more slack the system's covert requirement is, then more covert information can be transmitted, which makes the covert rate increase. Moreover, the covert rate increases as the transmit power of the interfering signals $P_{J}$ increases. That's why the more interference to Willie, its detection performance decreases, and we can take this opportunity to transmit the covert information, and thus our covert rate will increase.

\section{Conclusion}
In this proposed paper, we construct a model for covert communication using untrustworthy UAV relaying with the assistance of a full-duplex receiver, where we use interference from the full-duplex receiver to confuse the detection of the warden, and security constraints to limit eavesdropping by untrustworthy UAVs. First, we derive an expression for the probability of detection error on the warden, which in turn yields the optimal detection threshold to minimise the probability of detection error. Then, we derive expressions for the secure covert rate and the secure transmission rate, which in turn maximise the secure covert rate subject to satisfying the covert constraints and security constraints. Finally, we use extensive numerical results to verify the effectiveness of our scheme.

{10}

\begin{IEEEbiography}[{\includegraphics[width=1in,height=1.25in,clip,keepaspectratio]{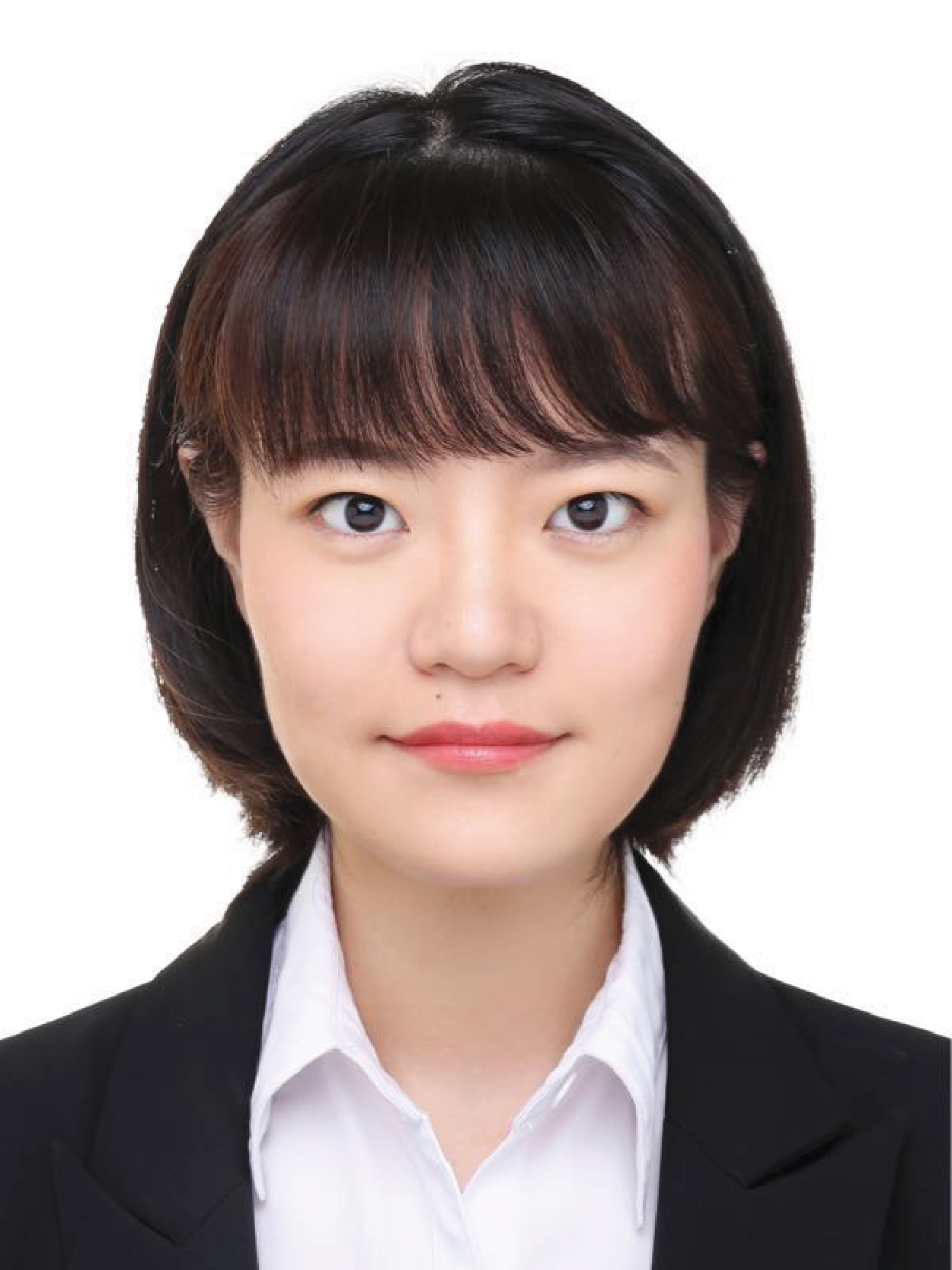}}]{Chan Gao}received her B.S. and M.S. degrees in Xi'an University of Posts and Telecommunications, Xi'an, China, in 2014 and 2018, and Ph.D. degree in systems information science from Future University Hakodate, Japan in 2021, respectively. She is currently a assistant professor at Xi'an University of Posts and Telecommunications and is also connected with the National Engineering Laboratory for Wireless Security, Xi'an, China. Her research interest focuses on the covert communication in physical layer.
\end{IEEEbiography}
\begin{IEEEbiography}[{\includegraphics[width=1in,height=1.25in,clip,keepaspectratio]{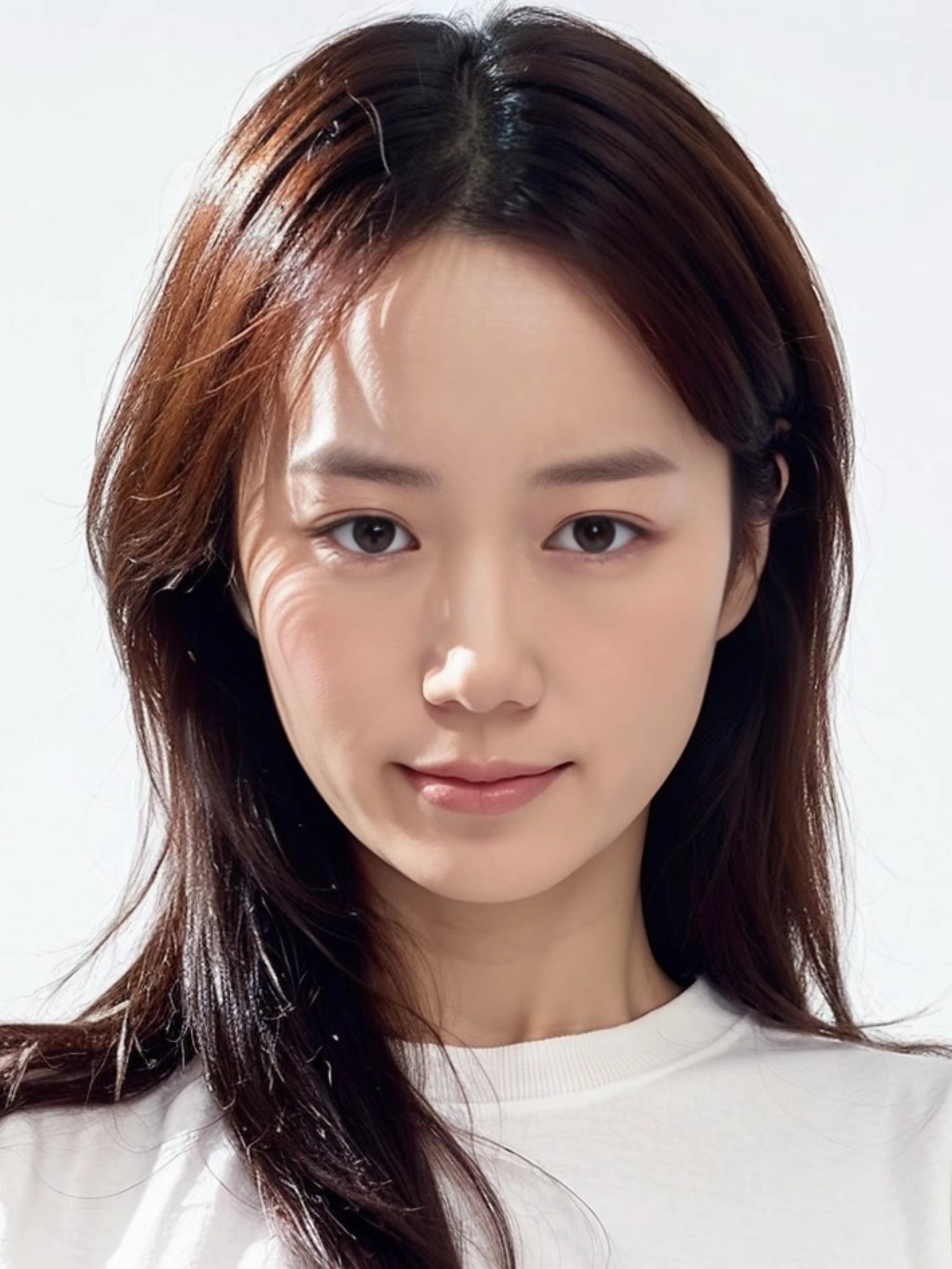}}]{Linying Tian}
Linying Tian received her B.S. degree from Xi'an University of Posts {\&}  Telecommunications in 2023. From 2023, She continues to pursue her M.S. degree in network and information security from Xi'an University of Posts {\&} Telecommunications. Her research interest focuses on the covert communication in physical layer.
\end{IEEEbiography}
\begin{IEEEbiography}[{\includegraphics[width=1in,height=1.25in,clip,keepaspectratio]{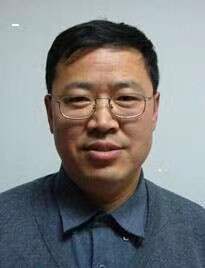}}]{Dong Zheng}received an M.S. degree in mathematics from Shaanxi Normal University, Xi'an, China, in 1988, and a Ph.D. degree in communication engineering from Xidian University,in 1999. He was a professor in the School of Information Security Engineering, Shanghai Jiao Tong University. He is currently a professor at Xi'an University of Posts and Telecommunications and is also connected with the National Engineering Laboratory for Wireless Security, Xi'an,China. He has published over 100 research articles including CT-RSA, IEEE Transactions on Industrial Electronics, Information Sciences. His research interests include cloud computing security, public key cryptography, and wireless network security.
\end{IEEEbiography}

\end{document}